\documentclass[letter,12pt]{article}%
\usepackage{amsmath,bm}
\usepackage{amsfonts}
\usepackage{amssymb}
\usepackage{graphicx}
\usepackage{rotating}
\usepackage[height=9in,left=1in,right=0.75in,bottom=1in]{geometry}
\usepackage[breaklinks=true,bookmarksopen=true,colorlinks=true,citecolor=blue]%
{hyperref}
\usepackage[authoryear]{natbib}
\usepackage{color}
\usepackage{colortbl}
\usepackage{paralist}
\usepackage{amsmath}%
\providecommand{\U}[1]{\protect\rule{.1in}{.1in}}
\RequirePackage{hypernat}

\catcode`~=11
\newcommand{\urltilde}{\kern -.15em\lower .7ex\hbox{~}\kern .04em}
\catcode`~=13
\def \@seccntformat#1{\csname the#1\endcsname.\quad}
\makeatother
\numberwithin{equation}{section}
\hypersetup{pdftitle={Escanciano}, pdfsubject={Robust Two-Step Semiparametric Estimation}, pdfauthor={Juan Carlos Escanciano}, pdfkeywords={Two-step estimation; Stochastic Equicontinuity;
Semiparametric inference} }
\begin{document}

\title{Nonseparable Multinomial Choice Models in Cross-Section and Panel Data}
\author{Victor Chernozhukov\thanks{Department of Economics, MIT, Cambridge, MA 02139,
U.S.A E-mail: vchern@mit.edu. }\\\textit{MIT}
\and Iv\'an Fern\'andez-Val\thanks{Department of Economics, Boston University,
Boston, MA 02215, U.S.A E-mail: ivanf@bu.edu. }\\\textit{Boston University}
\and Whitney K. Newey\thanks{Department of Economics, MIT, Cambridge, MA 02139,
U.S.A E-mail: wnewey@mit.edu. }\\\textit{MIT}}
\date{\today}
\maketitle

\begin{abstract}
\medskip

Multinomial choice models are fundamental for empirical modeling of economic
choices among discrete alternatives. We analyze identification of binary and
multinomial choice models when the choice utilities are nonseparable in
observed attributes and multidimensional unobserved heterogeneity with
cross-section and panel data. We show that derivatives of choice probabilities
with respect to continuous attributes are weighted averages of utility
derivatives in cross-section models with exogenous heterogeneity. In the
special case of random coefficient models with an independent additive effect,
we further characterize that the probability derivative at zero is
proportional to the population mean of the coefficients. We extend the
identification results to models with endogenous heterogeneity using either a
control function or panel data. In time stationary panel models with two
periods, we find that differences over time of derivatives of choice
probabilities identify utility derivatives ``on the diagonal,'' i.e. when the
observed attributes take the same values in the two periods. We also show that
time stationarity does not identify structural derivatives ``off the
diagonal'' both in continuous and multinomial choice panel models.

\medskip


\end{abstract}

\newpage

\section{Introduction}

Multinomial choice models are fundamental for empirical modeling of economic
choices among discrete alternatives. Our starting point is the assumption that
much of what determines preferences is unobserved to the econometrician. This
assumption is consistent with many empirical demand and other studies where
prices, income, and other observed variables explain only a small fraction of
the variation in the data. From the beginning unobserved preference
heterogeneity has had an important role in multinomial choice models. The
classic formulation of McFadden (1974) allowed for unobserved heterogeneity
through an additive term in the utility of each alternative. Hausman and Wise
(1978) developed a more general specification where coefficients of regressors
vary in unobserved ways among agents. Our results build on this pioneering
work as well as other contributions to be discussed in what follows.

Economic theory does not generally restrict the way unobserved heterogeneity
affects preferences. This observation motivates allowing for general forms of
heterogeneity, as we do in this paper. We allow choice utilities to depend on
observed characteristics and unobserved heterogeneity in general ways that
need not be additively or multiplicatively separable. The specifications we
consider allow for random coefficients but also more general specifications.

In this paper we show that derivatives of choice probabilities with respect to
continuous observed attributes are weighted averages of utility derivatives.
These results allow us to identify signs of utility derivatives as well as
relative utility effects for different attributes. We also find that
probability derivatives can be even more informative in special cases, such as
random coefficients. For example, we find that for linear random coefficients
with an independent additive effect the probability derivative at zero is
proportional to the population mean of the coefficients.

We give choice probability derivative results for binary and multinomial
choice. We do this for cross-section data where unobserved heterogeneity is
independent of observed attributes. We also give derivative formulas for two
cases with endogeneity. One is where the heterogeneity and utility variables
are independent conditional on a control function. There we show that
derivatives of choice probabilities conditional on the control function have a
utility derivative interpretation. We also verify that under a common support
condition, averaging over the control function gives structural function derivatives.

We also allow for endogeneity by using panel data. We give derivative formulas
for discrete choice in panel data under the time stationarity condition of
Manski (1987). For the constant coefficient case these give new identification
results for ratios of coefficients of continuously distributed variables in
panel data without requiring infinite support for any regressor or
disturbance. The panel data results are partly based on Hoderlein and White
(2012) as extended to the time stationary case by Chernozhukov et al. (2015).
These results use the "diagonal" where regressors in two time periods are
equal to each other.

We also consider identification "off the diagonal," where regressors in
different time periods are not equal to each other. For the case of a single
regressor and two time periods we construct an alternative, observationally
equivalent model that is linear in the regressor. This alternative model can
have a different average utility derivative off the diagonal, showing that
utility average derivatives are not identified there.

The model and goal of this paper are different than that of Burda, Harding,
and Hausman (2008, 2010) and Gautier and Kitamura (2013) Their goal is recover
the distribution of heterogeneity in a linear random coefficients model. We
consider a more general nonseparable model and a more modest goal of obtaining
weighted average effects from probability derivatives. Our results provide a
way of recovering certain averages of utility derivatives. Also, our results
are simpler in only depending on nonparametric regressions rather than the
Bayesian or deconvolution methods required to identify distributions of random coefficients.

Section 2 gives derivative formulae for binary choice. Section 3 extends these
results to multinomial choice models. Section 4 obtains derivative results in
the presence of a control function. Section 5 gives identification results for
multinomial choice in panel data. Section 6 shows nonidentification off the
diagonal. Section 7 concludes. The Appendix gives proofs.

\bigskip

\section{Binary Choice Model}

We first consider a binary choice model in cross-section data where we observe
$(Y_{i},X_{i}),(i=1,...,n)$ with $Y\in\{0,1\}$ a binary choice variable and
$X$ a vector of observed characteristics (regressors). Let $\varepsilon$ be a
vector that is possibly infinite dimensional, representing unobserved aspects
of agents' preferences. We will assume that the utility of choices $0$ and $1$
is given by $U_{0}(X,\varepsilon)$ and $U_{1}(X,\varepsilon)$ respectively.
The binary choice variable $Y$ is%
\[
Y=1(U_{1}(X,\varepsilon)\geq U_{0}(X,\varepsilon)). \label{binary choice}%
\]
Here we impose no restrictions on the way that $X$ and $\varepsilon$ interact.
As we will discuss, this specification includes but is not limited to random
coefficient models. This specification is general enough to be like the
stochastic revealed preference setting of McFadden and Richter (1991).

We begin our analysis under the assumption that $X$ and $\varepsilon$ are
independently distributed:

\bigskip

\textsc{Assumption 1:} \textit{(Independence)} $X$\textit{ and }$\varepsilon
$\textit{ are independently distributed.}

\bigskip

In what follows we will relax this condition when we have a control function
or when we have panel data. It is helpful to think about this model as a
threshold crossing model where
\[
Y=1(\delta(X,\varepsilon)\geq0),\ \ \delta(X,\varepsilon)=U_{1}(X,\varepsilon
)-U_{0}(X,\varepsilon).
\]
The classic constant coefficients model is a special case where $\varepsilon$
is a scalar and
\[
\delta(X,\varepsilon)=\beta_{0}^{\prime}X+\varepsilon.
\]
This model only allows for additive unobserved heterogeneity. An important
generalization is a random coefficients model where $\varepsilon
=(v,\eta^{\prime})^{\prime}$ is a vector and
\[
\delta(X,\varepsilon)=\eta^{\prime}X+v.
\]
This specification allows for the coefficients of the regressors to vary with
the individual. Hausman and Wise (1978) proposed such a specification for
Gaussian $\varepsilon$. Berry (1994)  proposed a mixed logit/Gaussian
specification where $v$ is the difference of Type I extreme value variables
plus a constant and $\eta$ is Gaussian. Gautier and Kitamura (2013) gave
results on identification and estimation of the distribution of $\eta$ when
that distribution is unknown. The nonseparable specification we consider is
more general in allowing for $\delta(X,\varepsilon)$ to be nonlinear in $X$
and/or $\varepsilon$.

In this binary choice setting the choice probability is given by%
\[
P(X):=\Pr(Y=1\mid X)=\Pr(\delta(X,\varepsilon)\geq0\mid X)=\int1(\delta
(X,\varepsilon)\geq0)F_{\varepsilon}(d\varepsilon),
\]
where $F_{\varepsilon}$ denotes the CDF of $\varepsilon$. Here we derive a
formula that relates the derivatives of the choice probability with respect to
$X$ to the derivatives of $\delta(X,\varepsilon).$ Let $\partial_{x}$ denote
the vector of partial derivatives with respect to all the continuously
distributed components of $X$ and $\partial_{v}$ the partial derivative with
respect to a scalar $v.$

\bigskip

\textsc{Assumption 2:} \textit{(Monotonicity) (i) For some }$\eta$\textit{ and
}$v$\textit{, }$\varepsilon=(\eta^{\prime},v)^{\prime}$\textit{ where }%
$v$\textit{ is a scalar, }$\delta(x,\varepsilon)=\delta(x,\eta,v)$\textit{ is
continuously differentiable in }$x$\textit{ and }$v$\textit{, and there is
}$C>0$\textit{ such that }$\partial_{v}\delta(x,\eta,v)\geq1/C$\textit{ and
}$\Vert\partial_{x}\delta(x,\eta,v)\Vert\leq C$\textit{ everywhere. (ii) The
variable }$v $\textit{ is continuously distributed conditional on }$\eta
$\textit{ with a conditional density }$f_{v}(v\mid\eta)$\textit{ that is
bounded and continuous in }$v$\textit{.}

\bigskip

As discussed below, for binary choice this condition will be equivalent to
$\delta(x,\varepsilon)$ being additive in $v$ that is continuously distributed
with a density satisfying the above condition. Let $f_{\delta(x,\varepsilon)}$
denote the density of $\delta(x,\varepsilon)$.

\bigskip

\textsc{Theorem 1:} \textit{If Assumptions 1 and 2 are satisfied then,}
\[
\partial_{x}P(x)=\mathrm{E}[\partial_{x}\delta(x,\varepsilon)\mid
\delta(x,\varepsilon)=0]\cdot f_{\delta(x,\varepsilon)}(0).
\]

\bigskip

Theorem 1 shows that derivatives of the choice probability are scalar
multiples of averages of the derivative $\partial_{x}\delta(x,\varepsilon)$
conditional on being at the zero threshold, i.e. conditional on being
indifferent between the two choices. Here the choice probability is one minus
the CDF of $\delta(x,\varepsilon)$ at zero, so that the choice probability
derivative is the negative of the CDF derivative at zero, i.e.%
\[
\partial_{x}P(x)=-\partial_{x}\Pr(\delta(x,\varepsilon)\leq y)|_{y=0}.
\]
The formula in Theorem 1 corresponds to the derivative of the CDF of a
nonseparable model derived in Blomquist et al. (2014), which builds on the
quantile derivative result of Hoderlein and Mammen (2007). The conclusion of
Theorem 1 is an important application of this formula to the choice
probability derivative in a nonseparable model.

Assumption 2 restricts our model somewhat relative to the stochastic revealed
preference model of McFadden and Richter (1991). It is possible to obtain
another informative derivative formula under regularity conditions like those
of Sasaki (2015) and Chernozhukov, Fernandez-Val, and Luo (2015), that are
different than Assumption 2. Those conditions lead to a more general formula
for $\partial_{x}P(x)$. That formula allows for multiple crossings of the
threshold $0$ while the monotonicity condition in Assumption 2 implies that
there is only one threshold crossing conditional on $\eta$. It is not clear
how restrictive Assumption 2 or the alternative conditions are relative to the
stochastic revealed preference setting of McFadden and Richter (1991). For
brevity we omit further discussion of this issue.

Another special case of a nonseparable model is an index model where
$\delta(x,\varepsilon)=h(\beta_{0}^{\prime}x,\varepsilon)$ for some constant
coefficients $\beta_{0}$. Here $P(X)=\tau(\beta_{0}^{\prime}X)$ for
$\tau(u)=\Pr(h(u,\varepsilon)\geq0)=\int1(h(u,\varepsilon)\geq0)F_{\varepsilon
}(d\varepsilon).$ This model results in a choice probability that depends on
$X$ only through the index $\beta_{0}^{\prime}X,$ similarly to Stoker (1986),
Ichimura (1993), and Ai (1997). By Theorem 1 it follows that%
\[
\tau_{u}(u)=\partial_{u}\tau(u)=\mathrm{E}[\partial_{u}h(u,\varepsilon)\mid
h(u,\varepsilon)=0] \cdot f_{h(u,\varepsilon)}(0).
\]
Differentiating with respect to the continuous components of $X$ gives the
well known index derivative formula,%
\[
\partial_{x}P(x)=\beta_{0} \cdot\tau_{u}(\beta_{0}^{\prime}x).
\]
Here the derivatives of the choice probability are scalar multiples of the
components of $\beta_{0}$.

There is an alternative version of Theorem 1 that provides further insight and
motivates our multinomial choice results that follow in Section 3.
%Under the monotonicity condition of Assumption 2 we can assume without loss of
%generality that $\delta(x,\eta,v)$ is monotonic increasing in $v$.
%Define
%$h(x,\eta)= - \left.  \delta^{-1}(x,\eta,r)\right\vert _{r=0}$ where the function
%inverse is with respect to the $v$ argument in $\delta(x,\eta,v).$
By the monotonicity condition of Assumption 2
\[
\delta(x,\eta,v)\geq0\iff h(x,\eta)+v\geq0,
\]
where $h(x,\eta):=-\left.  \delta^{-1}(x,\eta,r)\right\vert _{r=0}$ and the
function inverse is with respect to the $v$ argument in $\delta(x,\eta,v).$
Then the choice probability is
\begin{equation}
P(x)=\mathrm{E}[\Pr(v\geq-h(x,\eta)\mid\eta)]=\mathrm{E}[1-F_{v}%
(-h(x,\eta)\mid\eta)]=\int[1-F_{v}(-h(x,\eta)\mid\eta)]F_{\eta}(d\eta
),\label{bin choice prob}%
\end{equation}
where $F_{v}(v\mid\eta)$ is the conditional CDF of $v$ given $\eta$, and
$F_{\eta}(\eta)$ is the CDF of $\eta.$ Differentiating the expression of
$P(x)$ in \eqref{bin choice prob} with respect to $x$ gives the following result:

\bigskip

\textsc{Theorem 2: }\textit{If Assumptions 1 and 2 are satisfied then}%
\[
\partial_{x}P(x)=\mathrm{E}[\left\{  \partial_{x}h(x,\eta)\right\}
f_{v}(-h(x,\eta)\mid\eta)]=\int\left[  \partial_{x}h(x,\eta)\right]
f_{v}(-h(x,\eta)\mid\eta)dF_{\eta}(\eta).
\]

\bigskip

This formula is easier to interpret than the formula in Theorem 1. Here we
clearly see that the derivative of the choice probability is a weighted
average of the derivative $\partial_{x}h(x,\eta)$ where the weight is the
conditional pdf of $v$ given $\eta$ evaluated at $-h(x,\eta).$

It is well known that the binary choice model is observationally equivalent to
a threshold crossing model where $h(x,\eta)$ is nonrandom. Let $\tilde
{h}(x)=P(x),$ $\tilde{v}$ be distributed $U(0,1)$ independently of $X$, and
$\tilde{Y}=1(P(X)-\tilde{v}\geq0)$. Then%
\[
\Pr(\tilde{Y}=1|X)=\Pr(\tilde{v}\leq P(X)|X)=P(X).
\]
This feature of binary choice models is not important for our purposes. Our
purpose is to provide interpretations of $P(x)$ and its derivatives in the
case where choice utilities vary across individuals in more complicated ways
than through an additive effect. Essentially we have strong, a priori views
that the utilities of different individuals are not just additive shifts of
one another.

An important kind of varying utility is one with random coefficients, where
$h(x,\eta)=x^{\prime}\eta.$ In this case the conclusion of Theorem 2 is that%
\[
\partial_{x}P(x)=\mathrm{E}[f_{v}(-\eta^{\prime}x|\eta)\eta].
\]
It is interesting to note when $v$ is independent of $\eta$ that at $x=0$%
\begin{equation}
\left.  \partial_{x}P(x)\right\vert _{x=0}=\mathrm{E}[f_{v}(0)\eta
]=f_{v}(0)\cdot\mathrm{E}[\eta]. \label{deriv at zero}%
\end{equation}
Thus, when $X$ has positive density around zero and $v$ and $\eta$ are
independent, the derivative of the choice probability at zero estimates the
expected value of the random coefficients up to scale. Consequently%
\[
\left.  \left[  \partial_{x_{j}}P(x)/\partial_{x_{k}}P(x)\right]  \right\vert
_{x=0}=\mathrm{E}[\eta_{j}]/\mathrm{E}[\eta_{k}]\text{.}%
\]
This equation is a binary choice analog of the result that in a linear random
coefficients model the regression of $Y$ on $X$ estimates the expectation of
the coefficients. With binary choice only ratios of coefficients are
identified, so here only ratios of expected values are identified.

\bigskip

\textsc{Corollary 3:} \textit{If Assumptions 1 and 2 are satisfied, }%
$\delta(x,\varepsilon)=\eta^{\prime}x+v,$\textit{ and }$v$\textit{ is
independent of }$\eta,$\textit{ then equation (\ref{deriv at zero}) is
satisfied.}

\bigskip

Higher order derivatives of the choice probability are also informative about
the distribution of random coefficients. For example, when $v$ is independent
of $\eta$ then differentiating twice with respect to $x$ gives%
\[
\left.  \frac{\partial^{2}P(x)}{\partial x\partial x^{\prime}}\right\vert
_{x=0}=-E[\eta\eta^{\prime}]f_{vv}(0),
\]
where $f_{vv}(v) = \partial_v f_v(v)$. 
Thus the second derivative of the probability is the second moment matrix of
the random coefficients, up to scale. When $f_{vv}(0)$ is nonzero this result
allows us to identify correlations among random coefficients as well as
relative variances. It follows similarly that higher order derivatives will be
scalar multiples of higher-order moments of $\eta.$

Weighted average derivatives of the choice probability can be used to
summarize the effect of $x$ on $h(x,\eta)$. From Theorem 2 we can see that
weighted average derivatives will be weighted averages of $\partial
_{x}h(x,\eta)$ conditional on $v = -h(x,\eta)$. In particular, for
any bounded nonnegative function $w(x)$ it follows that
\[
\mathrm{E}[w(X)\partial_{x}P(X)]=\mathrm{E}[w(X)f_{v}(-h(X,\eta)\mid
\eta)\left\{  \partial_{x}h(X,\eta)\right\}  ].
\]
Here the derivative is weighted by both $w(X)$ and the density $f_{v}%
(-h(X,\eta)\mid\eta)$. The density weight is present because the derivatives
of $h(x,\eta)$ have been \textquotedblleft filtered" through the discrete
choice and so the probability derivative only recovers effects where
$h(x,\eta)+v=0$.

 An example that we will consider repeatedly is random coefficients logit. For
binary choice this model has $v=\xi+\rho,$ where $\xi$ is a constant and
$\rho$ is the difference of two Type I extreme value disturbances that are
independent of $\eta.$ Let $f_{v}(v)=e^{\xi-v}/[1+e^{\xi-v}]^{2}$ be the logistic
pdf with location $\xi$. Then the conclusion of Theorem 2 gives%
\[
\partial_x P(x) =E\left[\frac{e^{\xi+x^{\prime}\eta}}%
{(1+e^{\xi+x^{\prime}\eta})^{2}}\eta\right].
\]
Here the probability derivative is a weighted average of the random
coefficients, with the weight being the logistic pdf values evaluated at the
regression $\xi+\eta^{\prime}x.$

\section{Multinomial Choice Models}

In this Section we extend the analysis to the nonseparable multinomial choice
model. Here there are $J$ choices $j=1,...,J.$ Each choice has a utility
$U_{j}(X,\varepsilon)$ associated with it, depending on observed
characteristics $X$ and unobserved characteristics $\varepsilon$. Let $Y_{j}$
denote the choice indicator that is equal to one if the $j^{th}$ alternative
is chosen and zero otherwise. Then
\[
Y_{j}=1(\{U_{j}(X,\varepsilon)\geq U_{k}(X,\varepsilon);k=1,...,J\}).
\]
The probability $P_{j}(x):=\Pr(Y_{j}=1\mid X=x)$ that $j$ is chosen
conditional on $X=x$ is the probability that $U_{j}(x,\varepsilon)$ is the
maximum utility among the $J$ choices, i.e.
\begin{align*}
P_{j}(x) =\Pr\{U_{j}(x,\varepsilon)\geq U_{k}(x,\varepsilon);k=1,...,J\}
=\int1(\{U_{j}(x,\varepsilon)\geq U_{k}(x,\varepsilon
);k=1,...,J\})F_{\varepsilon}(d\varepsilon),
\end{align*}
where we maintain Assumption 1 and assume that the probability of ties is zero.

We can obtain a useful formula for the derivative of this probability under a
condition analogous to Assumption 2. Recall that the monotonicity condition of
Assumption 2 is equivalent to the existence of a scalar additive disturbance.
Here we will impose scalar additive disturbances from the outset.

\bigskip

\textsc{Assumption 3:} \textit{(Multinomial Choice) There are }$\eta
,v_{j},u_{j}(x,\eta),(j=1,...,J)$\textit{ such that }$\varepsilon
=(\eta^{\prime},v^{\prime})^{\prime}$\textit{ for }$v:=(v_{1},...,v_{J}),$%
\[
U_{j}(x,\varepsilon)=u_{j}(x,\eta)+v_{j},
\]
\textit{and} $u_{j}(x,\eta)$\textit{ is continuously differentiable in }%
$x$\textit{ with bounded derivative.}

\bigskip

In this condition we assume directly an additive disturbance condition that we
showed is equivalent to Assumption 2 in the binary case. Assumption 3
generalizes that additive disturbance condition to multinomial choice.
Similarly to binary choice, we are not sure what restrictions this additive
specification would impose in the stochastic revealed preference setting of
McFadden and Richter (1991). For brevity we do not give a result for a
multinomial version of Assumption 1 which is quite complicated.

As for binomial choice we could formulate the results in terms of differences
of utilities. However, we find it convenient to work directly with choice
specific utilities $U_{j}(x,\varepsilon)=u_{j}(x,\eta)+v_{j}$ rather than
differences. Let $u:=(u_{1},...,u_{J})$ denote a $J\times1$ vector of
constants and%
\[
p_{j}(u\mid\eta):=\Pr(u_{j}+v_{j}\geq u_{k}+v_{k};k=1,...,J\mid\eta).
\]
This $p_{j}(u\mid\eta)$ is just the usual multinomial choice probability,
conditioned on $\eta$. When the $f_{v}(v\mid\eta)$ is continuous, $p_{j}%
(u\mid\eta)$ will be continuously differentiable in each $u_{k}.$ Let
\[
p_{jk}(u\mid\eta):=\partial p_{j}(u\mid\eta)/\partial u_{k},\ \ u(x,\eta
):=(u_{1}(x,\eta),...,u_{J}(x,\eta))^{\prime}.
\]

\bigskip

\textsc{Theorem 4:} \textit{If Assumptions 1 and 3 are satisfied, the
conditional density }$f_{v}(v\mid\eta)$\textit{ of }$v$\textit{ given }$\eta
$\textit{ is continuous in }$v,$ and $p_{jk}(u\mid\eta),(j,k=1,...,J)$
\textit{are bounded, then }$P_{j}(x)$\textit{ is differentiable in }%
$x$\textit{ and}%
\[
\partial_{x}P_{j}(x)=\mathrm{E}\left[  \sum_{k=1}^{J}p_{jk}(u(x,\eta)\mid
\eta)\partial_{x}u_{k}(x,\eta)\right]  =\int\left[  \sum_{k=1}^{J}%
p_{jk}(u(x,\eta)\mid\eta)\partial_{x}u_{k}(x,\eta)\right]  F_{\eta}(d\eta).
\]

\bigskip

\textsc{Example 1:} 
% We will consider repeatedly the multinomial logit example. 
\textit{(Multinomial Logit Model) Here $v$ is a
vector of i.i.d. Type I\ extreme value random variables independent of $\eta$.
The conditional choice probabilities $p_{j}$ have the multinomial logit form%
\[
p_{j}(u\mid\eta)=\frac{e^{u_{j}}}{\sum_{k=1}^{J}e^{u_{k}}}.
\]
Define $\tilde{p}_{j}(x,\eta):=e^{u_{j}(x,\eta)}/\sum_{k=1}^{J}e^{u_{k}%
(x,\eta)}.$ Then,%
\[
\partial_{x}P_{j}(x)=\mathrm{E}\left[  \tilde{p}_{j}(x,\eta)\left\{
\partial_{x}u_{j}(x,\eta)-\sum_{k=1}^{J}\tilde{p}_{k}(x,\eta)\partial_{x}%
u_{k}(x,\eta)\right\}  \right]  .
\]
For example, if some $x^{\ell}$ affects only $u_{j_{\ell}}(x,\eta)$ for some
$j_{\ell}$, then%
\[
\partial_{x^{\ell}}P_{j}(x)=\int\tilde{p}_{j}(x,\eta)\left\{  1(j=j_{\ell
})-\tilde{p}_{j_{\ell}}(x,\eta)\right\}  \partial_{x^{\ell}}u_{j_{\ell}%
}(x,\eta)F_{\eta}(d\eta).
\]}

Another important class of examples are those where $u_{j}(x,\eta)=\eta^{\prime
}x^{j}+\xi_{j}$ for choice specific observable characteristics $x^{j}$ and
constant $\xi_{j}.$ This example is similar to Berry, Levinsohn and Pakes (1995) where $x^{j}$ could be
thought of as the characteristics of an object for choice $j$, such as
characteristics of the $j^{th}$ car type. Here an additional unit of some
component of $x^{j}$ affects the utility the same for each alternative $j$.
%i.e. there is no utility interaction between the choice $j$ and the
%characteristics $x^{j}$. [UNCLEAR: SEE REFEREE COMMENT 15]} 
In this class of examples,%
\[
\partial_{x^{k}}P_{j}(x)=\mathrm{E}[p_{jk}(u(x,\eta)\mid\eta)\eta].
\]
Here we see that the derivative of the $j^{th}$ choice probability with
respect to the regressor vector $x^{k}$ for the $k^{th}$ alternative is an
expectation of the random coefficients multiplied by a scalar $\partial
p_{j}(u(x,\eta)\mid\eta)/\partial u_{k}$. As in the binary case if $\eta$ is a
constant vector $\beta_{0}$ then%
\[
\partial_{x^{k}}P_{j}(x)=p_{jk}(u(x,\beta_{0}))\cdot\beta_{0},
\]
so that the derivative of the choice probability is proportional to $\beta
_{0}$ for all $x^{k}$. Also if $v$ is independent of $\eta$ so that
$p_{j}(u\mid\eta)=p_{j}(u)$, and each of the characteristic vectors is zero,
then the scalar is a constant and
\[
\left.  \partial_{x^{k}}P_{j}(x)\right\vert _{x^{1}=\cdots=x^{J}=0}=\left.
p_{jk}(u)\right\vert _{u=0}\cdot\mathrm{E}[\eta].
\]
Similarly to the binary case the derivative of the probability at the origin
is a scalar multiple of the expectation of the random coefficients.  Moreover, it can be shown as in the binary case that higher-order derivatives identify higher-order moments of $\eta$, up to scale.

\section{Control Functions}

A model where it is possible to allow for nonindependence between
$\varepsilon$ and $X$ is one where there is an observable or estimable control
function $w$ satisfying

\bigskip

\textsc{Assumption 4:} \textit{(Control Function)} $X$\textit{ and
}$\varepsilon$\textit{ are independently distributed conditional on }%
$w$\textit{.}

\bigskip

As shown in Blundell and Powell (2004) and Imbens and Newey (2009),
conditioning on a control function helps to identify objects of
interest.\footnote{Berry and Haile (2010) considered an alternative approach
based on the availability of ``special regressors'' and instrumental variables
satisfying completeness conditions in multinomial choice demand models where
the endogenous part of the unobserved heterogeneity is scalar. This approach
identifies the entire distribution of random utilities.} Here we show how a
control function can be used to estimate averages of utility derivatives.
These derivatives will be exactly analogous to those considered previously,
except that we also condition on the control function.

The choices $Y_{j}$ are determined as before but now we consider choice
probabilities that condition on $w$ as well as $X.$ These probabilities are
given by%
\[
P_{j}(X,w):=\Pr(Y_{j}=1\mid X,w).
\]
Let $u:=(u_{1},...,u_{J})$ denote a $J\times1$ vector of constants and%
\[
p_{j}(u\mid\eta,w):=\Pr(u_{j}+v_{j}\geq u_{k}+v_{k};k=1,...,J\mid\eta,w).
\]
This $p_{j}(u\mid\eta,w)$ is just the usual multinomial choice probability,
conditioned on $\eta$ and $w.$ When the conditional density of $v$ given
$\eta$ and $w$ is continuous, $p_{j}(u\mid\eta,w)$ will be continuously
differentiable in each $u_{k}.$ Let $p_{jk}(u\mid\eta,w):=\partial p_{j}%
(u\mid\eta,w)/\partial u_{k}$ and $u(x,\eta):=(u_{1}(x,\eta),...,u_{J}%
(x,\eta))^{\prime}$ as before.

\bigskip

\textsc{Theorem 5:} \textit{If Assumptions 3 and 4 are satisfied, the
conditional density }$f_{v}(v\mid\eta,w)$\textit{ of }$v$\textit{ given }%
$\eta$\textit{ and }$w$ \textit{is continuous in }$v,$ \textit{and }%
$p_{jk}(u\mid\eta,w),(j,k=1,...,J)$ \textit{are bounded,} \textit{then }%
$P_{j}(x,w)$\textit{ is differentiable in }$x$\textit{ and}%
\begin{multline*}
\partial_{x}P_{j}(x,w)=\mathrm{E}\left[  \sum_{k=1}^{J}p_{jk}(u(x,\eta
)\mid\eta,w)\partial_{x}u_{k}(x,\eta)\mid w\right] \\
=\int\left[  \sum_{k=1}^{J}p_{jk}(u(x,\eta)\mid\eta,w)\partial_{x}u_{k}%
(x,\eta)\right]  F_{\eta}(d\eta\mid w).
\end{multline*}

\bigskip

\textsc{Example 1 (cont.):} \textit{Consider the  multinomial logit  model where $v$ is also independent of $w.$
% Define $\tilde{p}_{j}(x,\eta):=e^{u_{j}(x,\eta)}/\sum_{k=1}^{J}e^{u_{k}(x,\eta)}.$ 
Then,
\begin{align*}
\partial_{x}P_{j}(x,w) &  =\mathrm{E}\left[  \tilde{p}_{j}(x,\eta)\left\{
\partial_{x}u_{j}(x,\eta)-\sum_{k=1}^{J}\tilde{p}_{k}(x,\eta)\partial_{x}%
u_{k}(x,\eta)\right\}  \mid w\right]  \\
&  =\int\tilde{p}_{j}(x,\eta)\left\{  \partial_{x}u_{j}(x,\eta)-\sum_{k=1}%
^{J}\tilde{p}_{k}(x,\eta)\partial_{x}u_{k}(x,\eta)\right\}  F(d\eta\mid w).
\end{align*}
For example, if some $x^{\ell}$ affects only $u_{j_{\ell}}(x,\eta)$ for some
$j_{\ell}$, then%
\[
\partial_{x^{\ell}}P_{j}(x,w)=\int\tilde{p}_{j}(x,\eta)\{1(j=j_{\ell}%
)-\tilde{p}_{j\ell}(x,\eta)\}\partial_{x^{\ell}}u_{j_{\ell}}(x,\eta
)F(d\eta\mid w).
\]}

We can also obtain result  for the random coefficient model analogous to the previous section conditional on the control variable. We do not present these results for the sake of brevity.

%Consider again a BLP type example where $u_{j}(x,\eta)=\eta^{\prime}x^{j}%
%+\xi_{j}$ for choice specific observable characteristics $x^{j}$. Here%
%\[
%\partial_{x^{k}}P_{j}(x,w)=\int[p_{jk}(u(x,\eta)\mid\eta,w)\cdot\eta
%]F(d\eta\mid w).
%\]
%Here we see that the derivative of the $j^{th}$ choice probability with
%respect to the vector $x^{k}$ for the $k^{th}$ alternative is an expectation
%of the random coefficients multiplied by a scalar $\partial p_{j}%
%(u(x,\eta)\mid\eta,w)/\partial u_{k}$. As above, if $\eta$ is a constant
%vector $\beta_{0}$, then%
%\[
%\partial_{x^{k}}P_{j}(x,w)=\beta_{0}\cdot p_{jk}(u(x,\beta_{0})\mid w),
%\]
%so that the derivative of the choice probability is proportional to $\beta
%_{0}$ for all $x$ and $w.$ Also, if $v$ is independent of $\eta$ and $w$ so
%that $p_{j}(u\mid\eta,w)=p_{j}(u)$, then at $x=0$ the scalar is a constant
%and
%\[
%\left.  \partial_{x^{k}}P_{j}(x,w)\right\vert _{x^{1}=\cdots=x^{J}=0}=\left.
%p_{jk}(u)\right\vert _{u=0}\cdot\mathrm{E}[\eta\mid w].
%\]
%Similarly to above, the derivative of the probability at the origin is a
%scalar multiple of the expectation of the random coefficients, here
%conditional on the control function.

As is known from the previous literature, integrating over the marginal
distribution of the control function gives probability derivatives identical
to those for $X$ and $\varepsilon$ independent, when a common support
condition is satisfied:

\bigskip

\textsc{Corollary 6:} \textit{If Assumptions 3 and 4 are satisfied, the
conditional density }$f_{v}(v\mid\eta,w)$\textit{ of }$v$\textit{ given }%
$\eta$\textit{ and }$w$ \textit{is continuous in }$v$ \textit{and bounded, and
the conditional support for }$w$ given $X=x$ \textit{equals the marginal
support for }$w$, \textit{then }$P_{j}(x,w)$\textit{ is differentiable in }%
$x$\textit{ and}%
\[
\int\partial_{x}P_{j}(x,w)F_{w}(dw)=\int\left[  \sum_{k=1}^{J}p_{jk}%
(u(x,\eta)\mid\eta)\partial_{x}u_{k}(x,\eta)\right]  F_{\eta}(d\eta),
\]
\textit{where $F_{w}(w)$ is the CDF of $w$.}

\bigskip

It is interesting to note that the common support condition is not needed for
identification of interesting effects. Averages of utility derivatives are
identified from probability derivatives, conditional on the control function,
as in Theorem 5. Also, because $\eta$ is independent of $X$ conditional on
$w$, averages over $\eta$ conditional on $X$ can be identified by integrating
the objects in Theorem 5 over the conditional distribution of $w$ given $X$.
This integration gives local average probability responses analogous to the
local average response given in Altonji and Matzkin (2005). In addition,
averaging over the joint distribution of $X$ and $w$ gives average derivatives
analogous to those considered by Imbens and Newey (2009). None of these
effects rely on the common support condition.

\bigskip

\section{Panel Data}

Panel data can also help us identify averages of utility derivatives when $X$
and $\varepsilon$ are not independent. Invariance over time of the
distribution of $\varepsilon$ conditional on the observed $X$ for all time
periods can allow us to identify utility derivative averages analogous to
those we have considered. This invariance over time of the distribution of
$\varepsilon$ conditional on regressors is the basis of previous panel
identification results by Manski (1987), Honore (1992), Abrevaya (2000),
Chernozhukov et al. (2013), Graham and Powell (2012), Chernozhukov et al.
(2015), and is an important hypothesis in Hoderlein and White (2012). Pakes
and Porter (2014) and Shi et al. (2017) have given identification results for
multinomial choice models under this condition. These papers allow for some
time effects while Evdokimov (2010) allowed for general time effects while
imposing independence and additivity among disturbances.

In this Section we consider panel binary and multinomial choice models. We focus on the case of two time periods. 
%It is straightforward to extend the results to more than two time periods. 
We start with the panel version of the general nonseparable binary choice model of Section 2. Here  $Y_{t} \in \{0,1\}$ is the binary choice variable and $X_{t}$ the vector of observed characteristics (regressors) at time $t$. We consider the threshold crossing model
\begin{equation}\label{eq:pb}
Y_t = 1(\delta(X_t, \varepsilon_t) \geq 0), \ \ (t=1,2),
\end{equation}
where $\delta(X_t, \varepsilon_t)$ represents the difference in utility between choices $0$ and $1$ at time $t$, and $\varepsilon_t$ is a vector of unobserved heterogeneity at time $t$ which include time variant and time invariant components such as individual effects.  The time stationarity of the difference in utility is important to our results
though it may be possible to relax that condition similarly to Chernozhukov et
al. (2015).

With panel data we can replace the assumption of independence of $\varepsilon_t$
and $X_t$ with the following time stationarity condition that is automatically satisfied by the time invariant components of $\varepsilon_t$:

\bigskip

\textsc{Assumption 5}: \textit{(Time Stationarity) The distribution of
}$\varepsilon_{t}$\textit{ given } $\mathbf{X:=}(X_{1},X_{2})$\textit{ does not
depend on }$t.$

\bigskip

To identify averages of utility derivatives we use the choice probability
conditional on the regressors for both time
periods, given by
\[
\Pr(Y_t = 1 \mid \mathbf{X}) = \mathrm{E}[Y_{t} \mid X_{1},X_{2}]  = \int 1(\delta(X_t, \varepsilon) \geq 0) F_{\varepsilon}(d \varepsilon \mid X_1, X_2),
\]
where  the CDF of $\varepsilon_t,$ $F_{\varepsilon}$, does not depend on $t$ by Assumption 5. Assume that $\delta(X_t, \varepsilon_t)$ and $F_{\varepsilon}(d \varepsilon \mid X_1,X_2)$ are differentiable in $X_t$. Then  an argument similar to Theorem 1 yields
\begin{multline*}
\partial_{X_t} \mathrm{E}[Y_{t} \mid X_{1},X_{2}] =  \mathrm{E}[\partial_{X_t} \delta(X_t,\varepsilon_t)\mid \mathbf{X},
\delta(X_t,\varepsilon_t)=0]\cdot f_{\delta(X_t,\varepsilon_t)}(0 \mid \mathbf{X}) \\ +  \int 1(\delta(X_t, \varepsilon) \geq 0) \partial_{X_t} F_{\varepsilon}(d \varepsilon \mid X_1, X_2).
\end{multline*}
The first term is a scalar multiple of the average utility derivative conditional on the regressors at both periods and on being indifferent between the two choices at time $t$.  The second term is heterogeneity bias coming from the dependence between  $X_t$ and $\varepsilon_t$. 

The next result shows that differences of derivatives of the choice probability identify up to a constant the average utility derivative on the diagonal where the regressors do not change over the two periods. As in Chernozhukov et al. (2015), time stationarity allows us to difference out the confounding effect of $X_t$ that acts through the correlation of $X_t$ with
$\varepsilon_t$. 

\bigskip

\textsc{Theorem 7:} \textit{If Assumptions 2(i) and 5 are satisfied,} $v_t$ \textit{is continuously distributed conditional on } $\eta_{t}$\textit{ and }$\mathbf{X}$ \textit{with a conditional density }$f_{v}(v\mid\eta_{t},\mathbf{X})$ \textit{that is bounded and continuous in
}$v,$  \textit{the conditional density} $f_{\eta}(\eta \mid X_1,X_2)$  \textit{ of} $\eta_t$ \textit{ given} $(X_1,X_2)$ \textit{is continuous in} $X_2$ \textit{and there is} $\delta >0$ \textit{such that} 
\begin{equation}\label{eq:th7}
\int \sup_{\| \Delta\| \leq \delta} f_{\eta}(\eta \mid X_1, X_2 + \Delta) d\eta < \infty,
\end{equation}
\textit{then,} 
$$
\partial_{X_2} \mathrm{E}[Y_2 - Y_1 \mid \mathbf{X}]\big|_{X_1 = X_2} =   \mathrm{E}[\partial_{X_2} \delta(X_2,\varepsilon_2)\mid \mathbf{X},
\delta(X_2,\varepsilon_2)=0] \big|_{X_1 = X_2} \cdot f_{\delta(X_2,\varepsilon_2)}(0 \mid \mathbf{X})\big|_{X_1 = X_2}
$$

\bigskip

We now turn to the multinomial choice model. Let
$Y_{jt}$ denote the choice indicator, equal to $1$ if alternative $j$ is
chosen in time period $t$. 
%Let $X_{t}$ denote the observable characteristics
%and $\varepsilon_{t}$ the unobservable ones in period $t.$ 
We assume that
choice is based on a time stationary utility function $U_{j}(x,\varepsilon
)=u_{j}(x,\eta)+v_{j}$ having the additive form considered in the previous
Section. Again the time stationarity of the utility may be possible to relax  similarly to Chernozhukov et
al. (2015).

It is assumed that the individual makes the choice that maximizes utility in
each time period, so that%
\[
Y_{jt}=1(\{U_{j}(X_{t},\varepsilon_{t})\geq U_{k}(X_{t},\varepsilon
_{t}),k=1,...,J\}),(j=1,...,J,t=1,2).
\]
To identify averages of utility derivatives we use again choice probabilities
conditional on the regressors  for both time
periods, given by
\[
P_{jt}(\mathbf{X}):=\Pr(Y_{jt}=1\mid X_{1},X_{2}).
\]
%With panel data we can replace the assumption of independence of $\varepsilon$
%and $\mathbf{X}$ with the following time stationarity condition:
%
%\bigskip
%
%\textsc{Assumption 5}: \textit{(Time Stationarity) The distribution of
%}$\varepsilon_{t}$\textit{ given } $\mathbf{X=}(X_{1},X_{2})$\textit{ does not
%depend on }$t.$
%
%\bigskip

For a constant vector $u:=(u_{1},...,u_{J})$ let%
\[
p_{j}(u\mid\eta_{t},\mathbf{X}):=\Pr(u_{j}+v_{tj}\geq u_{k}+v_{tk}%
;k=1,...,J\mid\eta_{t},\mathbf{X).}%
\]
This is like the usual choice probability, as discussed earlier, only now it
depends on $\mathbf{X}$ as well as $\eta_{t}.$ 
What allows us to identify
derivative effects despite the dependence of $p_{j}$ on $\mathbf{X}$ is that
$p_{j}$ does not depend on $t$ because of the time stationarity condition of
Assumption 5. Time stationarity allows us again to difference out the confounding
effect of $\mathbf{X}$ that acts through the correlation of $\mathbf{X}$ with
$\varepsilon$. By iterated expectations the choice probability is%
\[
P_{jt}(\mathbf{X})=\Pr(Y_{jt}=1\mid\mathbf{X)}=\mathrm{E}[p_{j}(u(X_{t}%
,\eta_{t})\mid\eta_{t},\mathbf{X})\mid\mathbf{X].}%
\]
The difference over two time periods is%
\[
P_{j2}(\mathbf{X})-P_{j1}(\mathbf{X})=\mathrm{E}[p_{j}(u(X_{2},\eta_{2}%
)\mid\eta_{2},\mathbf{X})\mid\mathbf{X]-}\mathrm{E}[p_{j}(u(X_{1},\eta
_{2})\mid\eta_{2},\mathbf{X)}\mid\mathbf{X],}%
\]
where we have used the time stationarity in replacing $\eta_{1}$ by $\eta_{2}$
in $P_{j1}(\mathbf{X}).$ When we differentiate this with respect to $X_{2}$
the presence of $P_{j1}(\mathbf{X})$ removes all the derivatives with respect
to $X_{2}$ except the utility derivatives, where $X_{1}=X_{2}.$

If the conditional density $f_{v}(v\mid\eta_{t},\mathbf{X)}$ is continuous in
$v$ then $p_{j}(u\mid\eta_{t},\mathbf{X})$ will be continuously differentiable
in $u.$ Let $p_{jk}(u\mid\eta_{t},\mathbf{X}):=\partial p_{j}(u\mid\eta
_{t},\mathbf{X})/\partial u_{k}$.
% and $\mu_{j}(x\mid\mathbf{X}):=\mathrm{E}%[p_{j}(u(x,\eta_{t})\mid\eta_{t},\mathbf{X})\mid\mathbf{X].}$

\bigskip

\textsc{Theorem 8: }\textit{If Assumptions 3 and 5 are satisfied, the
conditional density }$f_{v}(v\mid\eta_{t},\mathbf{X})$\textit{ of }$v_{t}%
$\textit{ given }$\eta_{t}$\textit{ and }$\mathbf{X}$ \textit{is continuous in
}$v,$ $p_{jk}(u\mid\eta_{t},\mathbf{X}),(j,k=1,...,J)$\textit{ are bounded,
and }$\mathrm{E}[p_{j}(u(x,\eta_{t})\mid\eta_{t},X_{1},X_{2})|X_{1}%
,X_{2}\mathbf{]}$\textit{ is differentiable in } $x$ \textit{and} $X_{2}%
$\textit{ then,}
\begin{align*}
\left.  \partial_{X_{2}}\mathrm{E}[Y_{j2}-Y_{j1}\mid\mathbf{X]}\right\vert
_{X_{1}=X_{2}} &  =\left.  \partial_{X_{2}}\{P_{j2}(\mathbf{X)-}%
P_{j1}(\mathbf{X)\}}\right\vert _{X_{1}=X_{2}}\\
&  =\left.  \mathrm{E}\left[  \sum_{k=1}^{J}p_{jk}(u(X_{2},\eta_{2})\mid
\eta_{2},\mathbf{X})\partial_{x}u_{k}(X_{2},\eta_{2})\mid\mathbf{X}\right]
\right\vert _{X_{1}=X_{2}}.
\end{align*}

\bigskip

\textsc{Example 1 (cont.):} \textit{Consider the multinomial logit where $v_{t}$ consists of i.i.d
Type I\ extreme value random variables that are independent of $\eta_{t}$ and
of $\mathbf{X.}$ 
%Define $\tilde{p}_{j}(x,\eta):=e^{u_{j}(x,\eta)}/\sum_{k=1}^{J}e^{u_{k}(x,\eta)}.$ 
Then,
\begin{multline*}
\left.  \partial_{X_{2}}\mathrm{E}[Y_{j2}-Y_{j1}\mid\mathbf{X]}\right\vert
_{X_{1}=X_{2}}\\
=\left.  \mathrm{E}\left[  \tilde{p}_{j}(X_{2},\eta_{2})\left\{  \partial
_{x}u_{j}(X_{2},\eta_{2})-\sum_{k=1}^{J}\tilde{p}_{k}(X_{2},\eta_{2}%
)\partial_{x}u_{k}(X_{2},\eta_{2})\right\}  \mid\mathbf{X}\right]  \right\vert
_{X_{1}=X_{2}}.
\end{multline*}
For example, if some $X_{2}^{\ell}$ affects only $u_{j_{\ell}}(X_{2},\eta)$,
then %
\[
\left.  \partial_{X_{2}^{\ell}}\mathrm{E}[Y_{j2}-Y_{j1}\mid\mathbf{X]}%
\right\vert _{X_{1}=X_{2}}=\left.  \mathrm{E}[\tilde{p}_{j}(X_{2},\eta
_{2})\{1(j=j_{\ell})-\tilde{p}_{j_{\ell}}(X_{2},\eta_{2})\}\partial_{x^{\ell}%
}u_{j_{\ell}}(X_{2},\eta_{2})\mid\mathbf{X}]\right\vert _{X_{1}=X_{2.}}.
\]}

An important class of examples is a panel version of Berry (1994) where $u_{j}%
(x,\eta)=\eta^{\prime}x^{j}$ for choice specific observable characteristics
$x^{j}$. In this class of examples,%
\[
\left.  \partial_{X_{2}^{k}}\mathrm{E}[Y_{j2}-Y_{j1}\mid\mathbf{X]}\right\vert
_{X_{1}=X_{2}}=\left.  \mathrm{E}[p_{jk}(u(X_{2},\eta_{2})\mid\eta
_{2},\mathbf{X})\cdot\eta_{2}\mid\mathbf{X}]\right\vert _{X_{1}=X_{2}}.
\]
Here we see that the derivative of the $j^{th}$ choice probability difference
with respect to $X_{2}^{k}$ for the $k^{th}$ alternative is an expectation of
the random coefficients multiplied by a scalar $p_{jk}(u(X_{2},\eta)\mid\eta
)$. The choice probabilities need not have the logit form for this result to hold. Also, analogous to the cross-section case, if $v_t$ is independent of $\eta_t$ conditional on $\mathbf{X}$ so that
$p_{j}(u\mid\eta
_{t},\mathbf{X})=p_{j}(u\mid \mathbf{X})$, and each of the characteristic vectors is zero at both time periods,
then the scalar is a constant and
\[
\left.  \partial_{X_{2}^{k}}\mathrm{E}[Y_{j2}-Y_{j1}\mid\mathbf{X]}\right\vert
_{X_{1}=X_{2} = 0} =\left.
p_{jk}(u\mid \mathbf{X}) \right\vert _{X_{1}=X_{2} =u=0}\cdot \left.\mathrm{E}[\eta \mid \mathbf{X}]\right\vert
_{X_{1}=X_{2} = 0}.
\]
Hence the derivative of the probability at the origin
is a scalar multiple of the expectation of the random coefficients conditional on the regressor being zero at both time periods. Again it can be shown as in the cross-section case that higher-order derivatives identify higher-order moments of $\eta$ conditional on $\mathbf{X}$, up to scale.

Time stationary panel data provides a way of controlling for endogeneity of
prices in imperfectly competitive markets where the price is one element of
$X_{t}$. The time stationarity condition of Assumption 5 allows for unobserved
features of preferences corresponding to $\varepsilon_{t}$ to be correlated
with $\mathbf{X}$ in unspecified ways, as long as that relationship is the
same for each time period. In particular, as mentioned earlier, components of
$\varepsilon$ that do not vary over time automatically satisfy this condition.
In this sense Assumption 5 is a very general condition for preferences that do
not vary over time. It can also be extended to settings where the dimension
$t$ corresponds to different markets or locations instead of time periods.

Similar to the cross-section case, if $\eta$ is a constant vector $\beta_{0}$
then%
\begin{equation}
\left.  \partial_{X_{2}^{k}}\mathrm{E}[Y_{j2}-Y_{j1}\mid\mathbf{X]}\right\vert
_{X_{1}=X_{2}}=\left.  p_{jk}(u(X_{2},\beta_{0})\mid\mathbf{X})\right\vert
_{X_{1}=X_{2}}\cdot\beta_{0}.\label{panel id}%
\end{equation}
Thus we find that that the derivative of the choice probability is
proportional to $\beta_{0}$ for all $X_{1}=X_{2}$ in a panel data multinomial
choice model where $u_{j}(x,\eta)=\beta_{0}^{\prime}x^{j}$.

\bigskip

\textsc{Theorem 9: }\textit{If Assumption 5 is satisfied, }$U_{j}%
(x,\varepsilon)=\beta_{0}^{\prime}x^{j}+v_{j},$\textit{ and }$\mathrm{E}%
[p_{j}(u(x,\eta_{t})\mid \eta_{t},X_{1},X_{2})|X_{1},X_{2}\mathbf{]}$ \textit{
is differentiable in }$x$ \textit{ and }$X_{2}$,\textit{ then for each }$j$
and $k$, equation \eqref{panel id} is satisfied.
%\[
%\left.  \partial_{X_{2}^{k}}\E[Y_{j2}-Y_{j1}\mid\mathbf{X]}\right\vert
%_{X_{1}=X_{2}}=\left.  \E[p_{jk}(u(X_{2},\eta_{2})\mid\eta_{2},\mathbf{X}%
%)\mid\mathbf{X]}\right\vert _{X_{1}=X_{2}}\cdot\beta_{0}.
%\]
\textit{Also, if }$\left.  \mathrm{E}[p_{jk}(u(X_{1},\eta_{t})\mid\eta
_{t},\mathbf{X})\mid\mathbf{X]}\right\vert _{X_{1}=X_{2}}\neq0$\textit{ for
some }$j,$\textit{ }$k,$\textit{ and }$X_{1}$, \textit{then }$\beta_{0}%
$\textit{ is identified up to scale.}

\bigskip

This gives an identification result for multinomial choice models in panel
data. It shows that the vector of coefficients of continuous regressors in a
multinomial choice model with additive fixed effect is identified up to scale
from the diagonal where $X_{1}=X_{2}.$ This identification result holds even
if $X_{t}$ is bounded, unlike that of Manski (1987). It can also hold even
with $v_{t}$ having bounded support, unlike that of Shi et al. (2017). In
independent work Chen and Wang (2017) has recently shown that in panel binary
choice the entire vector $\beta_{0}$ can be identified up to scale if just one
component of $X_{t}$ is continuously distributed.

\bigskip

\section{Nonidentification Off the Diagonal}

\bigskip

The panel data results show identification of utility derivatives on the
diagonal where $X_{1}=X_{2}.$ We can also show that off the diagonal, where
$X_{1}\neq X_{2},$ utility derivatives are not identified with two time
periods.  Specifically, off the diagonal one can obtain multiple values of
conditional expectations of utility derivatives from the same 
the data.

To provide intuition we first show nonidentification for the smooth case where%
\begin{equation}
Y_{t}=\phi(X_{t},\varepsilon_{t})\text{,}(t=1,2), \label{cont model}%
\end{equation}
$X_{t}$ is a scalar, $\phi(x,\varepsilon)$ is continuously differentiable in
$x$, and the distribution of $\varepsilon_{t}$ given $\mathbf{X}=(X_{1}%
,X_{2})^{\prime}$ is time stationary. Suppose that equation (\ref{cont model})
is true. We can construct an alternative, observationally equivalent
nonseparable model with time stationary disturbances as%
\begin{align*}
Y_{t}  &  =\tilde{\varepsilon}_{a}+\tilde{\varepsilon}_{b}X_{t}=\tilde{\phi
}(X_{t},\tilde{\varepsilon}),\ \ \tilde{\phi}\left(  x,\tilde{\varepsilon
}\right)  :=\tilde{\varepsilon}_{a}+\tilde{\varepsilon}_{b}x,\ \ \tilde
{\varepsilon}:=(\tilde{\varepsilon}_{a},\tilde{\varepsilon}_{b})^{\prime},\\
\tilde{\varepsilon}_{a}  &  :=Y_{1}-\tilde{\varepsilon}_{b}X_{1}%
,\ \ \tilde{\varepsilon}_{b}:=\left(  Y_{2}-Y_{1}\right)  /(X_{2}-X_{1}).
\end{align*}
By construction $\tilde{\varepsilon}$ does not vary with $t$, so that it is
time stationary. Also, $Y_{t}=\tilde{\phi}(X_{t},\tilde{\varepsilon})$ so that
the alternative model is observationally equivalent to the original one.
Furthermore, the expected value of $\tilde{\phi}_{x}\left(  X_{2}%
,\tilde{\varepsilon}\right)  :=\partial_x \tilde{\phi}(x,\tilde{\varepsilon
})\vert_{x=X_{2}}$ conditional on $\mathbf{X}$ is
\begin{equation}
\mathrm{E}[\tilde{\phi}_{x}\left(  X_{2},\tilde{\varepsilon}\right)
\mid\mathbf{X}]=\mathrm{E}[\tilde{\varepsilon}_{b}\mid\mathbf{X}%
]=\frac{\mathrm{E}[\phi(X_{2},\varepsilon_{2})-\phi(X_{1},\varepsilon_{2}%
)\mid\mathbf{X}]}{X_{2}-X_{1}.}. \label{alt deriv}%
\end{equation}
In contrast%
\begin{equation}
\mathrm{E}[\phi_{x}\left(  X_{2},\varepsilon_{t}\right)  \mid\mathbf{X}%
]=\left. \mathrm{E}\left[  \partial_x \phi(x,\varepsilon_{2})%
\mid\mathbf{X}\right] \right\vert
_{x=X_{2}} . \label{true deriv}%
\end{equation}
In general the expected derivative in equation (\ref{alt deriv}) will not
equal the expected derivative in equation (\ref{true deriv}) when
$\mathrm{E}[\phi(x,\varepsilon_{2})\mid\mathbf{X}]$ is nonlinear in $x$ over
the set where $X_{1}\neq X_{2}.$ Thus we have constructed an observationally
equivalent nonseparable model with $\mathrm{E}[\tilde{\phi}_{x}\left(
X_{2},\tilde{\varepsilon}\right)  \mid\mathbf{X}]\neq\mathrm{E}[\phi
_{x}\left(  X_{2},\varepsilon_{2}\right)  \mid\mathbf{X}],$ implying that
$\mathrm{E}[\phi_{x}\left(  X_{2},\varepsilon_{2}\right)  \mid\mathbf{X}]$ is
not identified, on the set where $X_{1}\neq X_{2}.$ The following is a precise
statement of this nonidentification result.

\bigskip

\textsc{Theorem 10: }\textit{If i) }$Y_{t}=\phi(X_{t},\varepsilon_{t}%
),$\textit{ }$\varepsilon_{t}$\textit{ is time stationary conditional on}
$\mathbf{X}$; \textit{ ii) }$\phi(x,\varepsilon)$\textit{ is continuously
differentiable in }$x$\textit{ with bounded derivative; and iii) }%
\[
\mathrm{E}[\phi(X_{2},\varepsilon_{2})-\phi(X_{1},\varepsilon_{2})-\phi
_{x}\left(  X_{2},\varepsilon_{2}\right)  (X_{2}-X_{1})\mid\mathbf{X}]\neq0
\]
\textit{ for }$X_{1}\neq X_{2},$\textit{ then }$\mathrm{E}[\phi_{x}\left(
X_{2},\varepsilon_{2}\right)  \mid\mathbf{X}]$\textit{ is not identified on
the set }$X_{1}\neq X_{2}.$

\bigskip

For example suppose $\phi(x,\varepsilon)$ is quadratic in $x$ with
$\varepsilon=(\varepsilon_{a},\varepsilon_{b},\varepsilon_{c})^{\prime}$ and
$\phi(x,\varepsilon)=\varepsilon_{a}+\varepsilon_{b}x+\varepsilon_{c}x^{2}.$
Then
\[
\mathrm{E}[\phi(X_{2},\varepsilon_{2})-\phi(X_{1},\varepsilon_{2})-\phi
_{x}\left(  X_{2},\varepsilon_{2}\right)  (X_{2}-X_{1})\mid\mathbf{X}%
]=-E[\varepsilon_{c2}\mid\mathbf{X](}X_{2}-X_{1})^{2}.
\]
Theorem 9 implies that in this quadratic model $\mathrm{E}[\phi_{x}\left(
X_{2},\varepsilon_{2}\right)  \mid\mathbf{X}]$ is not identified on $X_{1}\neq
X_{2}$ when $E[\varepsilon_{c2}\mid\mathbf{X]\neq0.}$

It is interesting that the form of the alternative, observationally equivalent
model $Y_{t}=\tilde{\varepsilon}_{a}+\tilde{\varepsilon}_{b}X_{t}$ is linear
in $X_{t}$. This is the model considered by Graham and Powell (2012).
Observational equivalence of this model to the true model means that it is
impossible to distinguish from the data a linear in $x$ model from a nonlinear
one, when there is one regressor and two time periods. Furthermore, the proof
given above also shows that the object estimated by the Graham and Powell
(2012) estimator will be the expected difference quotient
\[
\mathrm{E}[\tilde{\varepsilon}_{b}]=\mathrm{E}\left[  \frac{\phi
(X_{2},\varepsilon_{2})-\phi(X_{1},\varepsilon_{2})}{X_{2}-X_{1}}\right]  .
\]
This could be an interesting object. Of course one might also be interested in
the expected derivative on the diagonal given by $\left.  \partial_{X_{2}%
}\mathrm{E}[Y_{2}-Y_{1}\mid\mathbf{X}]\right\vert _{X_{1}=X_{2}}$; see
Hoderlein and White (2012) and Chernozhukov et al. (2015). It might be best to
report both kinds of effects in practice, given the impossibility of
distinguishing a linear from a nonlinear model when there is a scalar $X_{t}$
and two time periods.

We can give an analogous result for binary choice. Consider the binary choice
panel model in \eqref{eq:pb} where $Y_{t} = 1(\delta(X_t, \varepsilon_t) \geq 0)$ with scalar $X_t$. Suppose that this model satisfies Assumption 2.
As in the smooth case, we can construct an alternative, observationally equivalent
nonseparable model with time stationary disturbances as%
\begin{align*}
Y_{t}  &  = 1(\tilde{\delta
}(X_{t},\tilde{\varepsilon}) \geq 0),\ \ \tilde{\delta}\left(  x,\tilde{\varepsilon
}\right)  :=\tilde{\varepsilon}_{a}+\tilde{\varepsilon}_{b}x,\ \ \tilde
{\varepsilon}:=(\tilde{\varepsilon}_{a},\tilde{\varepsilon}_{b})^{\prime},\\
\tilde{\varepsilon}_{a}  &  :=\delta(X_1, \varepsilon_1)-\tilde{\varepsilon}_{b}X_{1}%
,\ \ \tilde{\varepsilon}_{b}:=\left(  \delta(X_2, \varepsilon_2)-\delta(X_1, \varepsilon_1)\right)  /(X_{2}-X_{1}).
\end{align*}
Note that this model  also satisfies Assumption 2 because $\tilde{\delta}\left(  x,\tilde{\varepsilon
}\right) $ is monotonic in  $\tilde{\varepsilon}_{a}$. The nonidentification result that we give here shows that the 
 object of interest in Theorem 7 is different for the two observationally equivalent models when $X_1 \neq X_2$. Thus,
\begin{multline*}
 \mathrm{E}[ \tilde \delta_x(X_2, \tilde \varepsilon)\mid \mathbf{X},
\tilde \delta(X_2,\tilde \varepsilon)=0]  =  \frac{ \mathrm{E}\left[\delta(X_2, \varepsilon_2)-\delta(X_1, \varepsilon_1)\mid \mathbf{X},
\delta(X_2,\varepsilon_2)=0\right]}{X_2-X_1}   \\ \neq  \mathrm{E}[ \delta_x(X_2,\varepsilon_2)\mid \mathbf{X},
 \delta(X_2,\varepsilon_2)=0] 
 \end{multline*}
in general. Here we use again the notation $g_x(X_2,\varepsilon_2) := \partial_x g(x,\varepsilon_2)\vert_{x=X_2}$ for $g = \delta, \tilde \delta$. The result then follows by
$$
f_{\tilde \delta(X_2,\tilde \varepsilon)}(0 \mid \mathbf{X}) = f_{\delta(X_2,\varepsilon_2)}(0 \mid \mathbf{X}).
$$
The following is a precise statement of this nonidentification result.

\bigskip

\textsc{Theorem 11: }\textit{Under the assumptions of Theorem 7 and}%
\[
\mathrm{E}[\delta(X_{1},\varepsilon_{2})+ \delta_x(X_2,\varepsilon_2) (X_{2}-X_{1})\mid\mathbf{X},  \delta(X_2,\varepsilon_2) = 0]\neq0
\]
\textit{ for }$X_{1}\neq X_{2},$\textit{ then }
$$ \mathrm{E}[\delta_x(X_2,\varepsilon_2)\mid \mathbf{X},
\delta(X_2,\varepsilon_2)=0]  \cdot f_{\delta(X_2,\varepsilon_2)}(0 \mid \mathbf{X})$$
\textit{ is not identified on
the set }$X_{1}\neq X_{2}.$

\bigskip

\section{Conclusion}

Jerry Hausman pioneered the introduction of flexible forms of unobserved
heterogeneity in structural economic models for multinomial choice. This paper
follows this tradition by considering identification of nonseparable
multinomial choice models with unobserved heterogeneity that is unrestricted
in both the dimension and its interaction with observed attributes. Some of
our results are local in nature. For example, we show that derivatives of
choice probabilities identify average utility derivatives only for marginal
units that are indifferent between two choices with cross-section data and for
units that have time invariant attributes with time stationary panel data. It
would be interesting to characterize minimal conditions that permit extending
the identification of average utility derivatives to larger populations. We
leave this extension to future work.

\bigskip

%\appendix

\section{Appendix: Proofs of Theorems}

\textbf{Proof of Theorem 1:} The proof is similar to the proof of Lemma 1 of
Chernozhukov et al. (2015). Let $F_{v}(v\mid\eta)=\int_{-\infty}^{v}f_{v
}(u\mid\eta)du$. Under Assumptions 1 and 2,
\begin{align*}
P(x)=\Pr(Y=1\mid X=x)  &  =\int\boldsymbol{1}\{\delta(x,\eta,v)\geq0\}F_{v
}(dv\mid\eta)F_{\eta}(d\eta)\\
&  =\int\boldsymbol{1}\{v\geq\delta^{-1}(x,\eta,0)\}F_{v}(dv \mid\eta)F_{\eta
}(d\eta)\\
&  =1-\int F_{v}(\delta^{-1}(x,\eta,0)\mid\eta)F_{\eta}(d\eta).
\end{align*}
Differentiating with respect to $x$
\[
\partial_{x}P(x)= - \int f_{v}(\delta^{-1}(x,\eta,0)\mid\eta)\partial
_{x}\delta^{-1}(x,\eta,0)F_{\eta}(d\eta),
\]
where the conditions of Assumption 2 allow us to differentiate under the
integral. Note that by the inverse and implicit function theorems,
\[
\partial_{x}\delta^{-1}(x,\eta,0)=-\left.  \frac{\partial_{x}\delta(x,\eta
,v)}{\partial_{v}\delta(x,\eta,v)}\right\vert _{\delta(x,\eta,v)=0}.
\]
Also, by a change of variable
\[
\left.  \frac{f_{v}(v\mid\eta)}{\partial_{v}\delta(x,\eta,v)}\right\vert
_{\delta(x,\eta,v)=0}=f_{\delta(x,\eta,v)}(0\mid\eta),
\]
where $f_{\delta(x,\eta,v)}(\cdot\mid\eta)$ is the conditional density of
$\delta(x,\eta,v)$ given $\eta$. Then substituting in gives%
\begin{align*}
\partial_{x}P(x)  &  =\int f_{\delta(x,\eta,v)}(0\mid\eta)\partial_{x}%
\delta(x,\eta,v)|_{\delta(x,\eta,v)=0}F_{\eta}(d\eta)\\
&  =\mathrm{E}[\partial_{x}\delta(x,\eta,v)\mid\delta(x,\eta,v)=0]\cdot
f_{\delta(x,\eta,v)}(0)\\
&  =\mathrm{E}[\partial_{x}\delta(x,\varepsilon)\mid\delta(x,\varepsilon
)=0]\cdot f_{\delta(x,\varepsilon)}(0),
\end{align*}
since
\begin{align*}
\mathrm{E}[\partial_{x}\delta(x,\eta,v)\mid\delta(x,\eta,v)=0]  &
=\int\partial_{x}\delta(x,\eta,\delta^{-1}(x,\eta,0))dF_{\eta}(\eta\mid
\delta(x,\eta,v)=0)\\
&  =\int\partial_{x}\delta(x,\eta,v)|_{\delta(x,\eta,v)=0}\frac{f_{\delta
(x,\eta,v) }(0\mid\eta)}{f_{\delta(x,\eta,v)}(0)}F_{\eta}(d\eta),
\end{align*}
by the Bayes rule. \textit{Q.E.D}.

\bigskip

\textbf{Proof of Theorem 2:} Given in text.

\bigskip

\textbf{Proof of Corollary 3:} Given in text.

\bigskip

\textbf{Proof of Theorem 4:} By iterated expectations,
\[
P_{j}(x)=\mathrm{E}[p_{j}(u(x,\eta)\mid\eta)]=\int p_{j}(u(x,\eta)\mid
\eta)F_{\eta}(d\eta).
\]
Also by Assumption 3 and the chain rule, $p_{j}(u(x,\eta)\mid\eta)$ is
continuously differentiable in $x$ with bounded derivative%
\[
\sum_{k=1}^{J}p_{jk}(u(x,\eta)\mid\eta)\partial_{x}u_{k}(x,\eta).
\]
Interchanging the order of differentiation and integration is then allowed,
and the conclusion follows. \textit{Q.E.D.}

\bigskip

\textbf{Proof of Theorem 5:} By iterated expectations and independence of $v$
and $x$ given $w$
\[
P_{j}(x,w)=\mathrm{E}[p_{j}(u(x,\eta)\mid\eta,w)\mid w]=\int p_{j}%
(u(x,\eta)\mid\eta,w)F_{\eta}(d\eta\mid w).
\]
Also, by $f(v\mid\eta,w)$\textit{ }continuous in $v$ and bounded and the chain
rule, $p_{j}(u(x,\eta)\mid\eta,w)$ is continuously differentiable in $x$ with
bounded derivative%
\[
\sum_{k=1}^{J}p_{jk}(u(x,\eta)\mid\eta,w)\partial_{x}u_{k}(x,\eta).
\]
Interchanging the order of differentiation and integration is then allowed,
and the conclusion follows. Q.E.D.

\bigskip

\textbf{Proof of Corollary 6:} Given in text.

\bigskip

\textbf{Proof of Theorem 7:} The proof is similar to the proof of  Theorem 2 of
Chernozhukov et al. (2015). Let $H(x, X_2 ) = \Pr(\delta(x,\varepsilon) \geq 0 \mid X_1,X_2)$. By the same argument as in the proof of Theorem 1 conditional on $(X_1,X_2),$ $H(x, X_2 ) $ is differentiable in $x$ with 
$$
\partial_{x} H(x, X_2 )  =  \mathrm{E}[\partial_{x} \delta(x,\varepsilon)\mid \mathbf{X},
\delta(x,\varepsilon)=0]  \cdot f_{\delta(x,\varepsilon)}(0 \mid \mathbf{X}).
$$ 
From \eqref{eq:th7}, $H(x, X_2)$ is also differentiable in $X_2$ with
$$
\partial_{X_2} H(x, X_2)  = \int 1(\delta(x,\varepsilon) \geq 0) \partial_{X_2}F_{\varepsilon}(d\varepsilon \mid  X_1,X_2). 
$$
The result follows by 
$$
\partial_{X_2} \mathrm{E}[Y_{t} \mid X_{1},X_{2}] = 1(t=2) \partial_{x}  H(x,X_2)\big|_{x=X_t}   + \partial_{X_2} H(x, X_2) \big|_{x=X_t}.
$$
taking differences with $t=1$ and $t=2$, and evaluating at $X_1 = X_2$.
\textit{Q.E.D}.

\bigskip

\textbf{Proof of Theorem 8:} By iterated expectations,
\[
P_{jt}(\mathbf{X})=\mathrm{E}[p_{j}(u(X_{t},\eta_{t})\mid\eta_{t}%
,\mathbf{X})]=\int p_{j}(u(X_{t},\eta)\mid\eta,\mathbf{X})F_{\eta}(d\eta
\mid\mathbf{X}),
\]
where $F(\eta\mid\mathbf{X})$ denotes the CDF of $\eta_{t}$ conditional on
$\mathbf{X}$. Also by Assumption 3 and the chain rule $p_{j}(u(x,\eta)\mid
\eta,\mathbf{X})$ is continuously differentiable in $x$ with bounded
derivative and
\[
\partial_{x}p_{j}(u(x,\eta)\mid\eta,\mathbf{X})=\sum_{k=1}^{J}p_{jk}%
(u(x,\eta)\mid\eta,\mathbf{X})\partial_{x}u_{k}(x,\eta).
\]
It follows by the previous equation that the order of differentiating an
integration can be interchanged to obtain
\[
\partial_{x}\mu_{j}(x\mid\mathbf{X)=}\mathrm{E}\left[  \sum_{k=1}^{J}%
p_{jk}(u(x,\eta_{t})\mid\eta_{t},\mathbf{X})\partial_{x}u_{k}(x,\eta_{t}%
)\mid\mathbf{X}\right]  .
\]
Note that $P_{jt}(\mathbf{X)=}\mu_{j}(X_{t}\mid\mathbf{X)}$. Let $\mu
_{j}(x\mid\mathbf{X}):=\mathrm{E}[p_{j}(u(x,\eta_{t})\mid\eta_{t}%
,\mathbf{X})\mid\mathbf{X].}$Then by the chain rule we have%
\begin{align*}
\left.  \partial_{X_{2}}\mathrm{E}[Y_{j2}-Y_{j1}\mid\mathbf{X]}\right\vert
_{X_{1}=X_{2}} &  =\left.  \partial_{X_{2}}\{\mu_{j}(X_{2}\mid\mathbf{X}%
)-\mu_{j}(X_{1}\mid\mathbf{X})\}\right\vert _{X_{1}=X_{2}}\\
&  =\left.  \left.  \partial_{x}\mu_{j}(x\mid\mathbf{X)}\right\vert _{x=X_{2}%
}\right\vert _{X_{1}=X_{2}}+\left.  \left.  \partial_{X_{2}}\mu_{j}%
(x\mid\mathbf{X)}\right\vert _{x=X_{2}}\right\vert _{X_{1}=X_{2}}\\
&  -\left.  \left.  \partial_{X_{2}}\mu_{j}(x\mid\mathbf{X)}\right\vert
_{x=X_{1}}\right\vert _{X_{1}=X_{2}}\\
&  =\left.  \left.  \partial_{x}\mu_{j}(x\mid\mathbf{X)}\right\vert _{x=X_{2}%
}\right\vert _{X_{1}=X_{2}}\\
&  =\left.  \mathrm{E}\left[  \sum_{k=1}^{J}p_{jk}(u(X_{2},\eta_{2})\mid
\eta_{2},\mathbf{X})\partial_{x}u_{k}(X_{2},\eta_{2})\mid\mathbf{X}\right]
\right\vert _{X_{1}=X_{2}}.
\end{align*}
Interchanging the order of differentiation and integration is then allowed,
and the conclusion follows. Q.E.D.

\bigskip

\textbf{Proof of Theorem 9:} Given in text.

\bigskip

\textbf{Proof of Theorem 10:} Given in text.

\bigskip

\textbf{Proof of Theorem 11:} Given in text.

\bigskip

\section*{Acknowledgements}

We thank the editor, referee, and participants at Cambridge-INET and Cemmap
Panel Data Workshop for comments and Siyi Luo for capable research assistance.
We gratefully acknowledge research support from the NSF. We appreciate the
hospitality of the Cowles Foundation where much of the work for this paper was accomplished.

\bigskip

%\newpage

\setlength{\parindent}{-.5cm} \setlength{\parskip}{.1cm}

\begin{center}
\textbf{REFERENCES}
\end{center}

\textsc{Abrevaya, J. }(2000):\ "Rank Estimation of a Generalized Fixed-Effects
Regression Model," \textit{Journal of Econometrics} 95, 1-23.

\textsc{Ai, C.} (1997): \textquotedblleft A Semiparametric Maximum Likelihood
Estimator\textquotedblright, \textit{Econometrica} 65, 933-963.

\textsc{Altonji, J., and R. Matzkin} (2005): \textquotedblleft Cross Section
and Panel Data Estimators for Nonseparable Models with Endogenous
Regressors\textquotedblright, \textit{Econometrica} 73, 1053-1102.

\textsc{Berry, S. }(1994): "Estimating Discrete Choice Models of Product
Differentiation," \textit{Rand Journal of Economics} 25, 242-262.

\textsc{Berry, S., P. Haile }(2010): "Nonparametric Identification of
Multinomial Choice Demand Models with Heterogeneous Consumers," Cowles
Foundation Discussion Paper 1718.

\textsc{Berry, S., J. Levinsohn, A. Pakes }(1995): "Automobile Prices in
Market Equilibrium," \textit{Econometrica} 63, 841-890.

\textsc{Blundell, R.W. and J.L. Powell} (2004):\ "Endogeneity in
Semiparametric Binary Response Models," Review of Economic Studies 71, 655--679.

\textsc{Blomquist, S., A. Kumar, C.-Y. Liang, W.K. Newey} (2014): "Individual
Heterogeneity, Nonlinear Budget Sets, and Taxable Income," CEMMAP working
paper CWP21/14

\textsc{Burda, M., M. Harding, and J.A. Hausman} (2008): "A Bayesian Mixed
Logit-Profit Model for Multinomial Choice," \textit{Journal of Econometrics}
147, 232-246.

\textsc{Burda, M., M. Harding, and J.A. Hausman} (2010): "A Poisson Mixture
Model of Discrete Choice," \textit{Journal of Econometrics} 166, 184-203.

%\textsc{Chamberlain, G.} (2010): "Binary Response Models for Panel Data:
%Identification and Information," \textit{Econometrica}, 78: 159-168.

\textsc{Chen, S., and X. Wang} (2017), ``Semiparametric Estimation of a Panel
Data Model without Monotonicity or Separability,'' unpublished manuscript,
Hong Kong University of Science and Technology.

\textsc{Chernozhukov, V., Fernandez-Val, I., Hahn, J., and W. K. Newey }
(2013), \textquotedblleft Average and Quantile Effects in Nonseparable Panel
Models,\textquotedblright\ \emph{Econometrica}, 81(2), pp. 535--580.

\textsc{Chernozhukov, V., I. Fernandez-Val, S. Hoderlein, H. Holzman, and W.K.
Newey} (2015): "Nonparametric Identification in Panels Using Quantiles,"
\textit{Journal of Econometrics }188, 378--392.

\textsc{Chernozhukov, V., I. Fernandez-Val, Y. Luo} (2015):\textquotedblleft
The Sorted Effects Method: Discovering Heterogenous Effects Beyond Their
Averages,\textquotedblright\ working paper.

%\textsc{Chernozhukov, V., D. Chetverikov, M. Demirer, E. Duflo, C. Hansen, W.
%Newey, J. Robins }(2017): "Double/Debiased Machine Learning for Treatment and
%Structural Parameters," MIT working paper.

\textsc{Evdokimov, K.} (2010), ``Identification and Estimation of a
Nonparametric Panel Data Model with Unobserved Heterogeneity,'' unpublished
manuscript, Princeton University.

%\textsc{Evdokimov, K. }(2011): "Nonparametric Identification of a Nonlinear
%Panel Model with Application to Duration Analysis with Multiple Spells,"
%working paper, Princeton University.

\textsc{Gautier, E. and Y. Kitamura }(2013): "Nonparametric Estimation in
Random Coefficients Binary Choice Models," \textit{Econometrica} 81, Pages 581--607.

\textsc{Graham, B.W. and J.L. Powell }(2012), \textquotedblleft Identification
and Estimation of Average Partial Effects in ``Irregular'' Correlated Random
Coefficient Panel Data Models,\textquotedblright\ \textit{Econometrica} 80
(5), pp. 2105--2152.

\textsc{Hausman, J.A., and D. Wise} (1978): "A Conditional Probit Model for
Qualitative Choice: Discrete Decisions Recognizing Interdependence and
Heterogeneous Preferences," \textit{Econometrica} 46, 403-26.

\textsc{Hausman, J.A., and W.K. Newey} (2016): "Individual Heterogeneity and
Average Welfare," \textit{Econometrica} 84, 1225-1248.

\textsc{Hoderlein, S.,\ and E.\ Mammen} (2007): \textquotedblleft
Identification of Marginal Effects in Nonseparable Models without
Monotonicity,\textquotedblright\ \textit{Econometrica}, 75, 1513 - 1519.

%\textsc{S. Hoderlein, S., E. Mammen, and K. Yu.} (2011): "Non-parametric
%Models in Binary Choice Fixed Effects Panel Data," \textit{The Econometrics
%Journal} 14, 351--367.

\textsc{Hoderlein, S. and H. White,} (2012), ``Nonparametric identi cation in
nonseparable panel data models with generalized fixed effects,''
\textit{Journal of Econometrics} 168, 300-314.

\textsc{Honore, B.E.} (1992): "Trimmed Lad and Least Squares Estimation of
Truncated and Censored Regression Models with Fixed Effects,"
\textit{Econometrica} 60, 533-565

\textsc{Ichimura, H.} (1993): "Semiparametric Least Squares (SLS) and Weighted
SLS Estimation of Single-Index Models," \textit{Journal of Econometrics} 58, 71-120.

\textsc{Imbens, G. and W.K. Newey }(2009): \ "Identification and Estimation of
Triangular Simultaneous Equations Models Without Additivity,"
\textit{Econometrica} 77, 1481-1512.

\textsc{Manski, C.F. }(1987): "Semiparametric Analysis of Random Effects
Linear Models from Binary Panel Data," \textit{Econometrica} 55, 357-362.

\textsc{McFadden, D.} (1974): "Conditional Logit Analysis of Qualitative
Choice Behavior," in P. Zarembka (ed) \textit{Frontiers of Econometrics},
Academic Press, 105-142.

\textsc{McFadden, D.; K. Richter }(1991) "Stochastic Rationality and Revealed
Stochastic Preference," in J. Chipman, D. McFadden, K. Richter (eds)
\textit{Preferences, Uncertainty, and Rationality}, Westview Press, 161-186.

\textsc{Pakes, A. and J. Porter} (2014): "Moment Inequalities for
Semi-parametric Multinomial Choice with Fixed Effects," Working paper, Harvard University.

%\textsc{Powell, J.L., J.H. Stock, and T.M. Stoker }(1989): "Semiparametric
%Estimation of Index Coefficients," \textit{Econometrica} 57, 1403-1430.

\textsc{Sasaki, Y.} (2015): \textquotedblleft What Do Quantile Regressions
Identify for General Structural Functions?\textquotedblright%
\ \textit{EconometricTheory} 31, 1102-1116.

\textsc{Shi, X., Shum, M., and W. Song} (2017): "Estimating Semi-parametric
Panel Multinomial Choice Models using Cyclic Monotonicity," Working paper,
University of Wisconsin-Madison.

\textsc{Stoker, T. }(1986): "Consistent Estimation of Scaled Coefficients,"
\textit{Econometrica} 54, 1461-1482.

\setlength{\parindent}{.0cm} \setlength{\parskip}{.1cm}

\end{document}